\pdfoutput=1

\documentclass[twocolumn,aps,pra,superscriptaddress]{revtex4}

\usepackage{times}
\usepackage{graphicx}
\usepackage{epsfig}
\usepackage{amsfonts}
\usepackage{amsmath}
\usepackage{amssymb}
\usepackage{color}
\usepackage{multirow}
\usepackage[colorlinks=true,citecolor=blue,urlcolor=blue]{hyperref}

\def\endproof{\vrule height6pt width6pt depth0pt}

\begin{document}


\title{Quantum nonlocality via local contextuality with qubit-qubit entanglement}



%
\author{Debashis Saha}
\affiliation{Institute of Theoretical Physics and Astrophysics, University of Gda\'{n}sk, 80-952 Gda\'{n}sk, Poland}
\author{Ad\'{a}n~Cabello}
\affiliation{Departamento de F\'{\i}sica Aplicada II, Universidad de Sevilla, E-41012 Sevilla, Spain}
\author{Sujit K. Choudhary}
\affiliation{Institute of Physics, Sachivalaya Marg, Bhubaneswar-751005, Odisha, India}
\author{Marcin Paw\l{}owski}
\affiliation{Institute of Theoretical Physics and Astrophysics, University of Gda\'{n}sk, 80-952 Gda\'{n}sk, Poland}


\begin{abstract}
  Quantum nonlocality can be revealed ``via local contextuality'' in qudit-qudit entangled systems with $d > 2$, that is, through the violation of inequalities containing Alice-Bob correlations that admit a local description, and Alice-Alice correlations (between the results of sequences of measurements on Alice's subsystem) that admit a local (but contextual) description. A fundamental question to understand the respective roles of entanglement and local contextuality is whether nonlocality via local contextuality exists when the parties have only qubit-qubit entanglement. Here we respond affirmatively to this question. This result further clarifies the connection between contextuality and nonlocality and opens the door for observing nonlocality via local contextuality in actual experiments.
\end{abstract}




\maketitle

\section{Introduction}

The incompatibility between quantum theory (QT) and local hidden variable theories (LHVTs) was first established by Bell \cite{Bell,CHSH} by means of an inequality which involves correlations between
measurement results from two distantly located physical systems. This inequality must be satisfied by any LHVT, but there are quantum states and observables which violate it.

On the other hand, QT is also incompatible with noncontextual hidden variable theories (NCHVTs) \cite{Specker60,Bell66,KS67,Mermin93}. A hidden variable model
is said to be noncontextual if it assigns a value to any observable, independently of which other compatible observables are being measured. The contextuality (i.e., lack of noncontextuality) predicted by QT can be observed through the violation of inequalities which involve correlations among the outcomes of compatible measurements on the same system \cite{Cabello08,BBCP09,YO11}. The simplest quantum system which exhibits contextual behavior is a three-level quantum system (or qutrit).

The connection between quantum contextuality and quantum nonlocality was first discussed under the scope of the so-called ``Kochen-Specker with locality theorem'' \cite{HR83,Redhead87,BS90}, later called the ``free will theorem'' \cite{CK06,CK09}. Based on these ideas, in a recent work \cite{Cabello130}, one of the coauthors of this paper has shown that there is a fundamentally different way for revealing quantum nonlocality which does not involve classically inexplicable correlations between distant systems when Alice-Bob correlations and Alice-Alice correlations (i.e., correlations between successive measurements on one of the local systems) are examined separately, but it does when both types of correlations are considered together. Interestingly, the correlations between two distant systems can be reproduced by a local hidden variable model, but the correlations between successive measurements on the local systems cannot be reproduced by noncontextual models. This emphasizes the role of local contextuality in the nonlocality observed in these scenarios.

However, so far, all examples of ``quantum nonlocality via local contextuality'' \cite{Cabello130,Cabello12} require qudit-qudit entanglement with $d > 2$, in contrast with the minimum entanglement needed to show standard quantum nonlocality, namely, qubit-qubit entanglement. In other words, all scenarios considered earlier demand Schmidt rank of the quantum state to be at least three. This leads to the question of whether local contextuality can also play a role in nonlocality scenarios with qubit-qubit entanglement (where the shared entanglement is less than or equal to one). Since, contextuality requires local dimension to be greater than two, the answer to this question does not seem obvious.

In this paper, we answer this question affirmatively. For this purpose, we derive a Bell inequality from the assumption of LHVTs. This inequality has two parts. One part contains correlations among the results of successive measurements on Alice's subsystem. The other part contains correlations between the results of measurements on Alice's subsystem and spacelike separated results on Bob's subsystems. Then, we show that the predictions of QT violate this inequality when Alice and Bob share two qubits in a singlet state and Alice also has in her possession an ancillary qubit. Strikingly, the predictions of QT do not exceed the bound for LHVTs for the Alice-Bob correlations which means that Alice-Bob correlations admit a local description. Interestingly, when we consider also the Alice-Alice correlations, the whole set of correlations exhibits quantum nonlocality. As shown below, this nonlocality stems from the state-independent contextuality of Alice's subsystem. We, then, extend this study to
two qubits in nonmaximally entangled states and find that a large number of such states exhibit this feature. 
We also study this for three qubits (shared by Alice, Bob, and Charlie) in Greenberger-Horne-Zeilinger (GHZ) states \cite{ghz} as here also, the Schmidt rank is two for any bipartition of the system. The implications of these results are discussed at the end.


\section{Quantum nonlocality via local contextuality with qubit-qubit entanglement}


Consider a bipartite system in which Alice has two qubits (qubits 1 and 2) and Bob, who is spatially separated from Alice, has a third qubit (qubit 3). Alice can measure, on her two-qubit system, sequences of three compatible observables taken from the following table of $\pm 1$-valued observables \cite{Mermin93}:
\begin{align}
&A=z_{1},&B=z_{2},\;\;\;\;\;\;\;\;\;\;\;\;\;\;&C=z_{1} z_{2}, \nonumber \\
&a=x_{2},&b=x_{1},\;\;\;\;\;\;\;\;\;\;\;\;\;\;&c=x_{1} x_{2}, \nonumber \\
&\alpha=z_{1} x_{2},&\beta=x_{1} z_{2},\;\;\;\;\;\;\;\;\;\;\;\;\;\;&\gamma=y_{1} y_{2},
\label{observables}
\end{align}
where, e.g., $z_{1} x_{2}$ denotes $\sigma_z^{(1)} \otimes \sigma_x^{(2)}$, that
is, the tensor product of the Pauli matrices $Z$ of qubit 1 and
$X$ of qubit 2. Observables in the same row or column are compatible. Bob can measure two $\pm 1$-valued observables denoted as $P,Q$.

{\it Theorem 1:} Any LHVT satisfies the following inequality:
\begin{equation}\label{in1}
\begin{split}
\langle T \rangle + \langle S \rangle \stackrel{\mbox{\tiny{LHVT}}}{\leq} 18,
\end{split}
\end{equation}
where
\begin{equation}
\begin{split}\label{eq1}
\langle T \rangle = & \langle CAB \rangle + \langle BAC \rangle + \langle \alpha \beta \gamma \rangle \\& + \langle \beta \alpha \gamma \rangle + \langle a A \alpha \rangle + \langle \alpha A a \rangle + \langle B b \beta \rangle \\& + \langle \beta B b \rangle + \langle cab \rangle + \langle abc \rangle - \langle C c \gamma \rangle - \langle c C \gamma \rangle
\end{split}
\end{equation} and
\begin{equation}
\begin{split}\label{eq2}
\langle S \rangle = & {\langle ABP \rangle}_C + {\langle ACP \rangle}_B + {\langle \beta \gamma P \rangle}_\alpha + {\langle \alpha \gamma P \rangle}_\beta \\& + {\langle A \alpha P \rangle}_a - {\langle A a Q \rangle}_\alpha + {\langle b \beta Q \rangle}_B + {\langle B b Q \rangle}_\beta \\& - {\langle c \gamma Q \rangle}_C + {\langle C \gamma Q \rangle}_c + {\langle a b P \rangle}_c - {\langle b c Q \rangle}_a.
\end{split}
\end{equation} Here ${\langle ABP \rangle}_C$ denotes the correlation function $\langle ABP \rangle$ in those events
where $AB$ is measured in the sequential measurement $CAB$ on Alice's side.

{\it Proof:} Consider the following three Clauser-Horne-Shimony-Holt \cite{CHSH} Bell inequalities:
\begin{subequations}
\begin{align}
&\langle C P \rangle + \langle C Q \rangle + \langle \alpha P \rangle - \langle \alpha Q \rangle \stackrel{\mbox{\tiny{LHVT}}}{\leq} 2, \\
&\langle \beta P \rangle + \langle \beta Q \rangle + \langle c P \rangle - \langle c Q \rangle \stackrel{\mbox{\tiny{LHVT}}}{\leq} 2, \\
&\langle B P \rangle + \langle B Q \rangle + \langle a P \rangle - \langle a Q \rangle \stackrel{\mbox{\tiny{LHVT}}}{\leq} 2.
\end{align}
\end{subequations}
Adding all of them, we obtain
\begin{equation}\label{b}
\begin{split}
& \langle C P \rangle + \langle C Q \rangle + \langle \alpha P \rangle - \langle \alpha Q \rangle +
\langle \beta P \rangle + \langle \beta Q \rangle \\ & + \langle c P \rangle - \langle c Q \rangle +
\langle B P \rangle + \langle B Q \rangle + \langle a P \rangle - \langle a Q \rangle \stackrel{\mbox{\tiny{LHVT}}}{\leq} 6.
\end{split}
\end{equation}

Now recall that Alice is allowed to perform sequential measurements on her two-qubit system. We want to prove that some subsets of Alice-Bob correlations in inequality (\ref{b}) are lower bounded by Alice-Alice-Alice and Alice-Alice-Bob correlations, without assuming noncontextuality in Alice's side. For example, we want to prove that
\begin{equation}\label{r}
\begin{split}
& \langle C P \rangle + \langle B P \rangle \geq \\
& \langle CAB \rangle + \langle BAC \rangle + {\langle ABP \rangle}_{C} + {\langle ACP \rangle}_B - 2.
\end{split}
\end{equation}
For that, let us denote by $\hat{O}$ the value assigned by the LHVT to observable $O$ when no other observable is measured first. This allows us to distinguish $\hat{O}$ from the values $O$ may have when other observables are measured before. Then, consider the following sequence of inequalities that hold due to simple algebraic constraints:
\begin{equation}\label{alg}
\begin{split}
& |\langle \hat{C} P \rangle + \langle \hat{B} P \rangle - \langle \hat{C}AB \rangle - \langle \hat{B}AC \rangle |
\\& \leq |\langle \hat{C} P \rangle - \langle \hat{C}AB \rangle| + |\langle \hat{B} P \rangle - \langle \hat{B}AC \rangle|
\\& \leq \langle |\hat{C}P - \hat{C}AB| \rangle + \langle |\hat{B}P - \hat{B}AC| \rangle
\\& \leq \langle |\hat{C}P - \hat{C}ABP^2| \rangle + \langle |\hat{B}P - \hat{B}ACP^2| \rangle
\\& = \langle |\hat{C}P(1 - ABP)| \rangle + \langle |\hat{B}P(1 - ACP)| \rangle
\\& = \langle |\hat{C}P||(1 - ABP)| \rangle + \langle |\hat{B}P||(1 - ACP)| \rangle
\\& = 1 - {\langle ABP \rangle}_C + 1 - {\langle ACP \rangle}_B.
\end{split}
\end{equation}
By comparing the first and last terms in (\ref{alg}) and replacing $\hat{C}AB$ by $CAB$ by assuring that $C$ is always measured in the first place, and replacing $\hat{B}AC$ by $BAC$, we obtain inequality (\ref{r}), where the notation does not preclude noncontextuality in Alice's side.

Similarly, we can probe that
\begin{equation}\label{r1}
\begin{split}
\langle \alpha P \rangle + \langle \beta P \rangle \geq \langle \alpha \beta \gamma \rangle + \langle \beta \alpha \gamma \rangle + {\langle \beta \gamma P \rangle}_\alpha + {\langle \alpha \gamma P \rangle}_\beta - 2, \\
\langle a P \rangle - \langle \alpha Q \rangle \geq \langle a A \alpha \rangle + \langle \alpha A a \rangle + {\langle A \alpha P \rangle}_a - {\langle A a Q \rangle}_\alpha - 2, \\
\langle B Q \rangle + \langle \beta Q \rangle \geq \langle B b \beta \rangle + \langle \beta B b \rangle + {\langle b \beta Q \rangle}_B + {\langle B b Q \rangle}_\beta - 2, \\
 \langle C Q \rangle - \langle c Q \rangle \geq - \langle C c \gamma \rangle - \langle c C \gamma \rangle - {\langle c \gamma Q \rangle}_C + {\langle C \gamma Q \rangle}_c - 2, \\
\langle c P \rangle - \langle a Q \rangle \geq \langle cab \rangle + \langle abc \rangle + {\langle a b P \rangle}_c - {\langle b c Q \rangle}_a - 2.
\end{split}
\end{equation}
Substituting Eqs.~(\ref{r}) and (\ref{r1}) in Eq.~(\ref{b}), we obtain inequality (\ref{in1}). \hfill \endproof

The term $\langle S \rangle$ defined in Eq.~(\ref{eq2}) contains all the Alice-Bob correlations. The upper bound of $\langle S \rangle$ for locally noncontextual LHVTs [denoted as (LC)LHVTs] can be obtained by assigning all possible deterministic values of the observables. To obtain the upper bound of $\langle S \rangle$ for any LHVT, including those that are locally contextual (but locally nondisturbing), we can consider a particular situation where the observables $P,Q$ are $+1$ valued and the probability distribution of the sequential measurements (irrespective of the sequence) on Alice's subsystem is given here, 
\begin{center}
\begin{tabular}{ c|c|c|c|c|c|c } 
 & $ABC$ & $Aa\alpha$ & $abc$ & $Bb\beta$ & $\alpha \beta \gamma$ & $Cc\gamma$ \\
 \hline
$ p(+,+,+)$ & $\frac{1}{2}$ & 0 & 0 & $\frac{1}{2}$ & $\frac{1}{2}$ & 0 \\ 
 \hline
 $p(+,+,-)$ & 0 & 0 & $\frac{1}{2}$ & 0 & 0 & 0 \\ 
 \hline
 $p(+,-,+)$ & 0 & $\frac{1}{2}$ & 0 & 0 & 0 & $\frac{1}{2}$ \\ 
 \hline
 $p(+,-,-)$ & 0 & 0 & 0 & 0 & 0 & 0 \\ 
 \hline
 $p(-,+,+)$ & 0 & 0 & 0 & 0 & 0 & 0 \\ 
 \hline
 $p(-,+,-)$ & 0 & $\frac{1}{2}$ & 0 & 0 & 0 & $\frac{1}{2}$ \\ 
 \hline
 $p(-,-,+)$ & 0 & 0 & $\frac{1}{2}$ & 0 & 0 & 0 \\ 
 \hline
 $p(-,-,-)$ & $\frac{1}{2}$ & 0 & 0 & $\frac{1}{2}$ & $\frac{1}{2}$ & 0 \\ 
 \hline
\end{tabular}
\end{center}
One can check that the above probability distribution satisfies the no-disturbance principle, i.e., the overall outcome statistics of an observable in sequential measurement is independent of whether any other compatible observable is measured or not. 
Such local assignment of values for all observables gives the maximum algebraic value of $\langle S \rangle$. We obtain
\begin{equation}
\langle S \rangle \stackrel{\mbox{\tiny{(NC)LHVT}}}{\leq} 10 \stackrel{\mbox{\tiny{LHVT}}}{\leq} 12.
\end{equation}
Later we show that the quantum predictions has no contradiction with LHVT when the quantity $\langle S \rangle$ is concerned.
On the other hand, the term $\langle T \rangle$ defined in Eq.~(\ref{eq2}) only
contains Alice-Alice-Alice correlations among three successive measurements on Alice's subsystem. It can also be easily checked that, for any NCHVT \cite{Cabello08},
\begin{equation}
\langle T \rangle \stackrel{\mbox{\tiny{NCHVT}}}{\leq} 8.
\label{a}
\end{equation}
Notice that all the observables in each term in $\langle T \rangle $ are mutually compatible. Indeed, inequality (\ref{a}) is a state-independent noncontextuality inequality in Alice's side \cite{Cabello08}.

To show that inequality (\ref{in1}) gets violated in QT, we consider the following state
shared between Alice (who has qubits 1 and 2) and Bob (who has qubit 3):
\begin{equation}
\begin{split}
&|\Psi \rangle_{12|3} = |\chi \rangle_1 |\psi^- \rangle_{23},\\
&|\chi \rangle_1 = \cos(\pi/8)|0 \rangle_{1} + \sin(\pi/8)|1 \rangle_{1},\\
&|\psi^-\rangle_{23} = \frac{1}{\sqrt{2}} (|0\rangle_2 |1\rangle_3-|1\rangle_2 |0\rangle_3).
\end{split}
\end{equation}
The observables at Bob's side are chosen as
\begin{equation}
\begin{split}
P=-\frac{(z_{3}+x_{3})}{\sqrt{2}},Q=-\frac{(z_{3}-x_{3})}{\sqrt{2}}.
\end{split}
\end{equation}
The quantities appearing in inequality (\ref{in1}) thus take the following values:
\begin{equation}
\begin{split}
&\langle ABP \rangle = \langle \beta \gamma P \rangle = \langle \alpha \gamma P \rangle = - \langle A a Q \rangle \\
&= \langle B b Q \rangle = - \langle c \gamma Q \rangle = \langle C \gamma Q \rangle = \langle abP \rangle = \frac{1}{2},\\
&\langle ACP \rangle = \langle A \alpha P \rangle = \langle b \beta Q \rangle = -\langle b c Q \rangle = \frac{1}{\sqrt{2}}.
\end{split}
\end{equation}
Therefore, $\langle S \rangle = 4+2\sqrt{2} = 6.828 (< 12)$. Nonetheless, $\langle T \rangle + \langle S \rangle = 18.828$, violates inequality (\ref{in1}).
Hence these correlations cannot have a description in terms of LHVTs. However, the Alice-Alice-Bob correlations (represented by $\langle S \rangle$) show no contradiction with LHVTs. Therefore, we can conclude that the violation of inequality (\ref{in1}), which was derived assuming only LHVTs, is revealed from the contextual correlation observed in Alice's subsystem and local Alice-Alice-Bob correlation.  

\subsection*{ White noise and nonmaximal entanglement}

One may think that this only occurs in the case of perfect correlations and vanishes either when there is white noise in Alice-Bob singlet state or when Alice and Bob are not sharing a maximally entangled pair of qubits. However, in the presence of white noise (when the shared state between Alice and Bob is $\rho=v|\psi^-\rangle \langle\psi^-| + (1-v)\frac{\mathcal{I}}{4}$), a simple calculation shows that inequality (\ref{in1}) is violated whenever the visibility, $v$ is larger than $\frac{6}{4+2\sqrt{2}} \approx 0.8787$.

On the other hand, the quantum violation is not restricted to the perfect correlations characteristic of the maximally entangled state. To show that, we consider the following normalized nonmaximally entangled state:
\begin{equation}
\begin{split}
|\psi^-\rangle_{23} = d_1 |01\rangle - d_2|10\rangle
\end{split}
\end{equation}
for some real $d_1,d_2$, and the following observables in Bob's side:
\begin{subequations}
\begin{align}
 & P=-\cos(t) z_{3}-\sin(t) x_{3}, \\& Q=  \cos(t') z_{3}-\sin(t') x_{3},
\end{align}
\end{subequations}
where
\begin{equation}
 \cos(t)=\cos(t')= \frac{1}{\sqrt{1 + 4(d_1d_2)^2}},\;\;\;\; \sin(t)=- \sin(t').
\end{equation}
A simple calculation leads to $\langle S \rangle = \sqrt{1 + 4(d_1d_2)^2}(2 + 2\sqrt{2})$. Then, inequality (\ref{in1}) is violated by QT if
\begin{equation}
\begin{split}
\sqrt{1 + 4(d_1d_2)^2}(2 + 2\sqrt{2}) > 6, \\
|d_1d_2| > 0.369,
\end{split}
\end{equation}
a condition which is satisfied by a large number of two-qubit entangled states.


\subsection*{Nonlocality via local contextuality with GHZ entanglement}
Here we address the question of what happens if, instead of qubit-qubit entanglement, Alice, Bob, and Charlie share a three-qubit GHZ state with one qubit in Alice's side (who also has an ancillary qubit needed for performing compatible sequential measurements). Charlie measures two $\pm1$-valued observables $U,V$. Studying this scenario is interesting, since the Schmidt rank is 2 for any bipartition of the system.

Following the method described earlier, one obtains a similar inequality:
\begin{equation}\label{in2}
\begin{split}
\langle T \rangle + \langle S' \rangle \stackrel{\mbox{\tiny{LHVT}}}{\leq} 18,
\end{split}
\end{equation}
where $\langle T \rangle$ is defined in Eq.(\ref{eq1}) and
\begin{equation}
\begin{split}
\langle S' \rangle = & {\langle ABPV \rangle}_C + {\langle ACPV \rangle}_B + {\langle \beta \gamma PU \rangle}_\alpha + {\langle \alpha \gamma PV \rangle}_\beta \\& + {\langle A \alpha PU \rangle}_a - {\langle A a QV \rangle}_\alpha + {\langle b \beta QU \rangle}_B + {\langle B b QU \rangle}_\beta \\& - {\langle c \gamma QU \rangle}_C + {\langle C \gamma QV \rangle}_c + {\langle a b PU \rangle}_c - {\langle b c QV \rangle}_a.
\end{split}
\end{equation}
To outline the proof, we start with the Bell inequality,
\begin{equation}\label{b1}
\begin{split}
& \langle C PV \rangle + \langle C QU \rangle + \langle \alpha PU \rangle - \langle \alpha QV \rangle +
\langle \beta PV \rangle + \langle \beta QU \rangle + \\ & \langle c PU \rangle - \langle c QV \rangle +
\langle B PV \rangle + \langle B QU \rangle + \langle a PU \rangle - \langle a QV \rangle \stackrel{\mbox{\tiny{LHVT}}}{\leq} 6.
\end{split}
\end{equation}
Taking pairs of terms, we derive six algebraic relations [as we did in Eq.~(\ref{alg})], for example:
\begin{equation}\begin{split}
& \langle C PV \rangle + \langle B PV \rangle \geq \\& \langle CAB \rangle + \langle BAC \rangle + {\langle ABPV \rangle}_C +{\langle ACPV \rangle}_B - 2.
\end{split}\end{equation}
Subsequently, Eq.~(\ref{in2}) can be obtained by substituting these algebraic relations in Eq. (\ref{b1}). \hfill \endproof

The four-qubit state shared by Alice, Bob, and Charlie is given by,
\begin{equation}
\begin{split}
 |\Psi \rangle_{12|3|4} = & |\chi \rangle_1 |\psi \rangle_{234},\\
 |\chi \rangle_1 = & \cos(\pi/8) |0\rangle_1 + \sin(\pi/8) |1\rangle_1,\\
 |\psi\rangle_{234} = & \frac{1}{2\sqrt{2}} (|000\rangle_{234}+|001\rangle_{234}+|010\rangle_{234}-|011\rangle_{234} \\
 & +|100\rangle_{234}-|101\rangle_{234}-|110\rangle_{234} - |111\rangle_{234}),
\end{split}
\end{equation}
where $|\psi\rangle_{234}$ is a GHZ state. The observables are chosen as follows:
\begin{equation}
P=z_3,Q=x_3;\;U=z_4V=x_4.
\end{equation}
It can be seen that, $\langle S' \rangle = 4\sqrt{2} + 4 = 9.657 (<12),$ such that $\langle T \rangle + \langle S' \rangle = 21.657$, and the corresponding threshold visibility of the GHZ state is $0.6213$.\\


\section{Conclusions}


Nonlocality via local contextuality is an exceptional way for revealing nonlocality  which connects nonlocality and local contextuality in certain scenarios. The implication of such a quantum feature is to show the equivalence between two different manifestations of quantum nonlocality. These two are: nonlocal correlation between two distant parties, and contextual correlation in the sequential measurement of one's subsystem with correlation between distant parties admitting a local model. This equivalence is natural in the two scenarios studied in the literature, namely two-ququart \cite{Cabello130} and two-qutrit \cite{Cabello12} systems in maximally entangled states. Similarly, state-independent contextuality assisted by bipartite entanglement has been exploited to show fully nonlocal quantum correlations  \cite{CabelloAVN,Brassard,Cabello2012}. Also in these cases, the required entanglement has Schmidt rank greater than 2. Since, contextuality requires local dimension to be greater than 2, an interesting question is whether local contextuality may lead to nonlocality in the case of qubit-qubit entanglement (where the shared entanglement is less than or equal to 1). We have shown that the answer is affirmative. Then, we have shown that neither noise nor the lack of perfect entanglement impedes us to observe this effect. This makes this configuration valuable for observing quantum nonlocality via local contextuality in experiments, extending previous experiments without entanglement \cite{KZ09,BKS09,ARB09,LH09,MR09}, since it is easier to perform sequential measurements on two qubits such that just one of them is entangled with a distant location than requiring both qubits to be entangled with a distant location \cite{Cabello130} or requiring qutrit-qutrit entanglement \cite{Cabello12}.

We have also explored the role of the type of entanglement exploited in this form of nonlocality and shown that GHZ states shared by three parties, even noisy ones, also allow us to reveal quantum nonlocality via local contextuality.

Our examples highlight the variety of roles and uses of entanglement in nonlocality scenarios via local contextuality and suggest ways to avoid the so-called ``compatibility loophole'' \cite{KZ09,G10} typical of quantum contextuality experiments with sequential measurements. Nevertheless, an open question remains: Can any form of nonlocality be understood as based on local contextuality? More specifically, can quantum nonlocality via local contextuality be extended for all pure entangled states in bipartite and multipartite scenarios. This needs to be investigated in future research.


\begin{acknowledgments}
This work is supported by the FQXi large grant project ``The Nature of Information in Sequential Quantum Measurements,'' Project No.~FIS2014-60843 (MINECO, Spain) with FEDER funds, the NCN grant 2013/08/M/ST2/00626, and the IDSMM programme at University of Gda\'{n}sk. S.K.C. acknowledges support from the Council of Scientific and Industrial Research, Government of India (Scientists' Pool Scheme). D.S. and S.K.C. thank G.\ Kar for useful discussions.
\end{acknowledgments}





\end{document}